\documentclass[aps,prl,twocolumn,showpacs,superscriptaddress,letterpaper]{revtex4}

\usepackage[dvips,letterpaper,body={7in,9.6in}]{geometry}
\usepackage{graphicx,xspace}
\usepackage{txfonts}            
\usepackage{units}
\usepackage[breaklinks=true]{hyperref}         

\newcommand{\degC}{\ensuremath{\!^{\circ}C}}

\begin{document}

\title{Highly-ordered graphene for two dimensional electronics}
\author{J. Hass}
\affiliation{The Georgia Institute of Technology, Atlanta, Georgia
30332-0430, USA}
\author{C. A. Jeffrey}
\affiliation{Department of Physics and Astronomy, University of
Missouri-Columbia, Columbia, MO 65211}
\author{R. Feng}
\author{T. Li}
\author{X. Li}
\author{Z. Song}
\affiliation{The Georgia Institute of Technology, Atlanta, Georgia
30332-0430, USA}
\author{C. Berger}
\affiliation{CNRS-LEPES, BP166, 38042 Grenoble Cedex,
France}
\author{W. A. de Heer}
\author{P. N. First}
\author{E. H. Conrad}
\affiliation{The Georgia Institute of Technology, Atlanta, Georgia
30332-0430, USA}

\begin{abstract}
  With expanding interest in graphene-based electronics, it is crucial
  that high quality graphene films be grown.  Sublimation of Si from
  the 4H-SiC(0001) (Si-terminated) surface in ultrahigh vacuum is a
  demonstrated method to produce epitaxial graphene sheets on a
  semiconductor. In this paper we show that graphene
  grown from the SiC$(000\bar{1})$ (C-terminated) surface are of
  higher quality than those previously grown on SiC(0001).  Graphene grown on the C-face can have structural domain
  sizes more than three times larger than those grown on the Si-face
  while at the same time reducing SiC substrate disorder from
  sublimation by an order of magnitude.
\end{abstract}


\maketitle

An increasingly large effort is underway to create materials
suitable for nanometer-scale electronic devices.  One proposed
avenue is to take advantage of the unique electronic properties of
carbon nanotubes to make gates and ballistic conducting
wires.\cite{Avouris02} Challenges for such an approach are:
Control of the properties of individual nanotubes (e.g. diameter,
helicity), inherent heterojunction impedances associated with
interconnection of nanotubes, and the assembly of vast networks
from individual nanotube devices.  Similar challenges are inherent
to any approach that relies on preformed nanoscale objects.  A
more conventional means to achieve large-scale integration of
nanoelectronic devices would be to rely on the continued scaling
of lithographic techniques, which have been the semiconductor
industry's greatest feat.  Assuming such advances in lithography,
a key issue is then the choice of material for nano-patterning.  A
suitable material should have excellent transport properties (e.g.
mobility) and allow for control of electronic properties (band
gap, doping) down to nanometer sizes.

It has been proposed that the unique electronic properties of
carbon nanotubes could be obtained if graphene sheets were limited
to nanometer-scale dimensions. \cite{Nakada96} Recent experiments
have demonstrated the unique electronic properties of graphene
\cite{Novoselov04a,Novoselov05b,Zhang05c} thus charting a
potential route to nanoelectronics based on epitaxial graphene
(EG). \cite{Berger04} A requirement for further progress will be
the preparation of a controlled number (thickness) of very large
epitaxial graphene sheets on semiconductor substrates.  These can
be lithographically-patterned into narrow ribbons or other shapes,
providing the necessary confinement for devices. Thus a scalable
assembly of nano-patterned epitaxial graphene (NPEG) devices
should be possible, with ballistic graphene conductors as
interconnects.
\begin{figure*}[htbp]
\begin{center}
\includegraphics[width=16.0cm,clip]{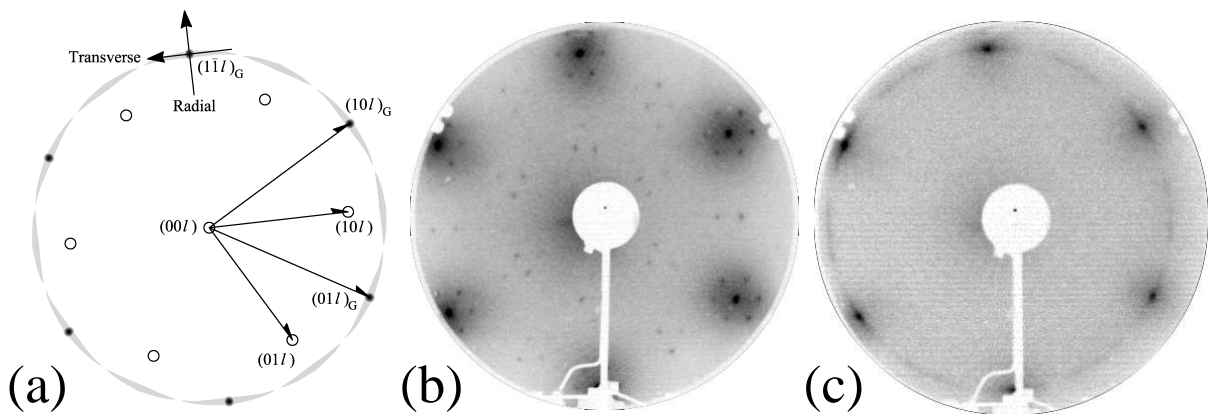}
\end{center}
\vspace{-1em}
\caption{(a) A schematic of reciprocal space for graphite on SiC.
Graphite rods ($\bullet$) are rotated $30^\circ$ from the SiC rods
($\circ$).  Shaded regions represent the effect of orientational
disorder in the graphite.  Radial and transverse x-ray scan
directions for Fig.~\ref{F:graphite_rocking} are depicted in (a).
(b) and (c) show LEED images acquired at \unit{75-eV} electron
energy: (b) a SiC(0001) Si-face surface with 2 graphene layers
(UHV synthesis); (c) SiC(000$\bar{1}$) C-face surface with 7
graphene layers (induction furnace synthesis).}
\label{F:LEED_image}
\end{figure*}

Prior investigations of 6H- and 4H- SiC(0001) and $(000\bar{1})$
surfaces showed that graphite films can be grown on these surfaces
by sublimating Si from SiC during heating above
$\sim\!\unit[1200]{\degC}$ in ultrahigh vacuum
(UHV).\cite{vanBommel75,Charrier02,Forbeaux00} These studies
showed that graphite grows epitaxially on the (0001) Si-terminated
(Si-face) surface of SiC, while graphite grown on the C-terminated
$(000\bar{1})$ (C-face) surface was rotationally disordered and
under some conditions formed nanocaps instead of a smooth
film.\cite{Kusunoki00} Consequently, the C-face was initially
overlooked as a potential substrate for graphene-based
electronics.

In this paper we report that the classification of the C-face graphite
as poorly ordered compared to the Si-face is incorrect. We instead
show that the mean structural domain size on C-face graphite is much
larger than on the Si-face, and that the inherent substrate roughness
from sublimation is dramatically suppressed compared to the best
previously-reported Si-face films. The improved structural order
correlates with recent magnetotransport measurements, which show
electron mobilities of over $\unit[1]{m^2/V\cdot s}$ and coherence
lengths exceeding $\unit[1]{\mu m}$ in graphite films prepared on the
C-face of SiC.\cite{Berger06} These observations have important
implications for the science and technology of graphene.

All substrates were 4H-SiC purchased from Cree, Inc. \cite{Cree}
Prior to graphitization the samples were ultrasonically cleaned in
acetone and ethanol, then hydrogen-etched at \unit[1600]{\degC}
for 30min. The $\text{H}_2$ etching was done in a vacuum
RF-induction furnace with a \unit[200]{ccm} flow of $5\%$
$\text{H}_2$/95\% Ar at $\unit[\sim\!1]{atm}$.  This process
removed all surface scratches and left a regularly stepped
surface, as characterized by AFM.  A typical step terrace width is
$\gtrsim\!\unit[1]{\mu m}$ with a step height of
$\unit[5]{\AA}$ (this corresponds to a $\sim\! 0.03^{\circ}$
miscut).

Si-face 4H-SiC(0001) samples were prepared in UHV
($P<\unit[1\times10^{-10}]{Torr}$) by electron-bombardment
heating. Substrates were first heated to \unit[1100]{\degC} for
\unit[6]{min}, then to \unit[1320]{\degC} for \unit[8]{min} to
remove surface contamination and form a well ordered
$\sqrt{3}\times\!\sqrt{3}$ reconstruction.  They were subsequently
heated to \unit[1400--1440]{\degC} for \unit[6--12]{min} to create
graphene films 1--2 layers thick, as determined by x-ray
reflectivity. C-face 4H-SiC$(000\bar{1})$ samples were heated to
\unit[1430]{\degC} for \unit[5--8]{min} in a vacuum RF-induction
furnace ($P = \unit[3\times10^{-5}]{\text{Torr}}$).  These
parameters produced graphitic films of 7--13 graphene layers.

The x-ray scattering experiments were performed at the Advanced
Photon Source, Argonne National Laboratory, on the 6ID-B-$\mu$CAT
beam line at $16.2~$keV photon energy.  The graphite film
thickness for all samples was determined by measuring the x-ray
intensity as function of $\ell$ along the graphite
$(1,\bar{1},\ell)_{G}$ rod.\cite{Charrier02} The notation
$(h,k,\ell)_G$ identifies a reciprocal-space point in units of the
graphite hexagonal reciprocal lattice basis vectors: $a^*_G
=\unit[2.9508]{\AA^{-1}}$ and $c^*_G = \unit[1.8829]{\AA^{-1}}$.
Unsubscripted reciprocal-space coordinates $(h,k,\ell)$ refer to
the substrate 4H-SiC hexagonal reciprocal lattice units: $a^* =
\unit[2.3552]{\AA^{-1}}$ and $c^* = \unit[0.6230]{\AA^{-1}}$.

A reciprocal space schematic for epitaxial graphene on SiC is shown in
Fig.~\ref{F:LEED_image}(a).  Open circles depict the $(1\times 1)$
low-energy electron diffraction (LEED) pattern from an unreconstructed
SiC surface, while the filled circles are the $(1\times 1)_{G}$
pattern for a graphene or graphite film with hexagonal unit cell
rotated azimuthally by $30^\circ$ ($R30^\circ$) relative to the SiC
$(1\times 1)$ cell. The diffuse ring through the $(1 \times 1)_{G}$
graphite spots in Fig.~\ref{F:LEED_image}(a) represents the $(1\times
1)$ LEED pattern from a graphene film with rotational disorder
relative to the SiC surface.

Figure~\ref{F:LEED_image}(b) shows the LEED pattern obtained from
a 2-graphene-layer film grown on the Si-face of SiC. In addition
to the graphite and SiC $(1\times 1)$ patterns, it shows a complex
$6\sqrt{3}\times 6\sqrt{3}R30^{\circ}$ reconstruction, which is
well known and has been studied extensively in the literature (see
e.g. Refs.~\onlinecite{Forbeaux00,Li06} and references therein).
The LEED pattern from a C-face sample with 7 graphene layers is
shown in Fig.~\ref{F:LEED_image}(c).  The LEED only shows the
6-fold graphite pattern because the thicker C-face film attenuates
the electron beam. The azimuthal streaks in
Fig.~\ref{F:LEED_image}(c) may indicate some rotational disorder
in the graphene sheets (as previously observed\cite{Forbeaux00}).

\begin{figure}[htbp]
\begin{center}
\includegraphics[width=8.0cm,clip]{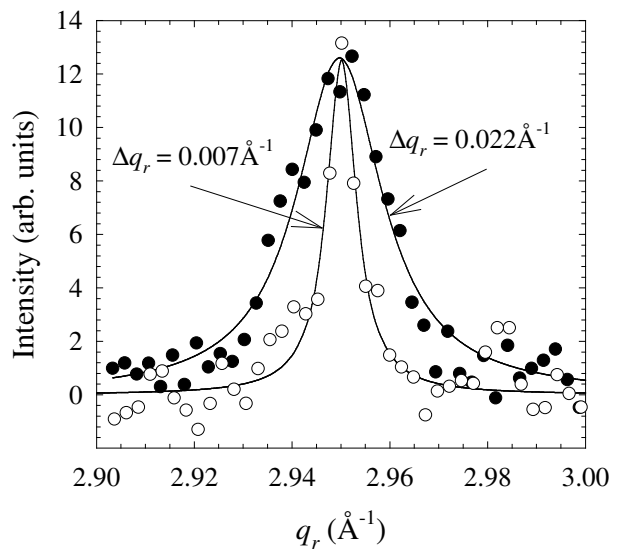}
\end{center}
\vspace{-1em}
\caption{Radial scans through the graphite crystal truncation rod
$(1,\bar{1},1.5)_G$ for both 2-layers of graphene grown on a
Si-face sample ($\bullet$) and 7-layers of graphene grown on a
C-face sample ($\circ$). Solid lines are Lorentzian fits.}
\label{F:graphite_rocking}
\end{figure}

More detailed information on the structural order of the samples was
acquired by surface x-ray scattering.  We have measured the width of
the graphite $(00\ell)_G$ and $(1\bar{1}\ell)_G$ rods on C- and
Si-face samples. Figure~\ref{F:graphite_rocking} shows radial scans
[see Fig.~\ref{F:LEED_image}(a)] across the graphite
$(1,\bar{1},1.5)_G$ crystal truncation rod for both the Si- and C-face
samples. The Si-face samples have a radial width of $\Delta q_{r}\sim
\unit[0.022]{\AA^{-1}}$ corresponding to a graphite mean coherent
domain size\cite{Lu82} $L=2\pi/\Delta q_{r}\sim \unit[290]{\AA}$. This
is very similar to the graphite domain size observed by Charrier {\it
  et al.}\cite{Charrier02} and is typical of the quality of graphite
grown on the SiC(0001) surface reported to date in the literature.
Although their surface treatment was different from ours, the fact
that the domain sizes are similar suggests a limit on the graphite
quality other than surface preparation.

In contrast, graphite grown on the C-face has much larger domain
sizes demonstrated by the smaller radial widths in
Fig.~\ref{F:graphite_rocking}: $\Delta q_{r}\sim
\unit[0.007]{\AA^{-1}}$, corresponding to
$L\sim\!\unit[900]{\AA}$. So while the LEED patterns show
azimuthal disorder in the C-face graphite, the coherent graphite
domains are more than 3 times larger than for the Si-face
graphite. This improved structural coherence correlates with the
high carrier mobility measured recently for C-face
graphene\cite{Berger06} versus Si-face.\cite{Berger04}  We note
that the difference in film thickness may play a role in the long
range order of the films.  However, the C-face $\Delta q_t$ do not
change for films between 7--13 layers.  There is also little
difference in the long range order of 1-2 layer graphene films on
the Si-face. In addition growth of 4--5 graphene layers on Si-face
requires temperature above $\sim\!\unit[1500]{\degC}$ where
substrate disorder becomes problematic (see below).

\begin{figure}[t]
\begin{center}
\includegraphics[width=8.0cm,clip]{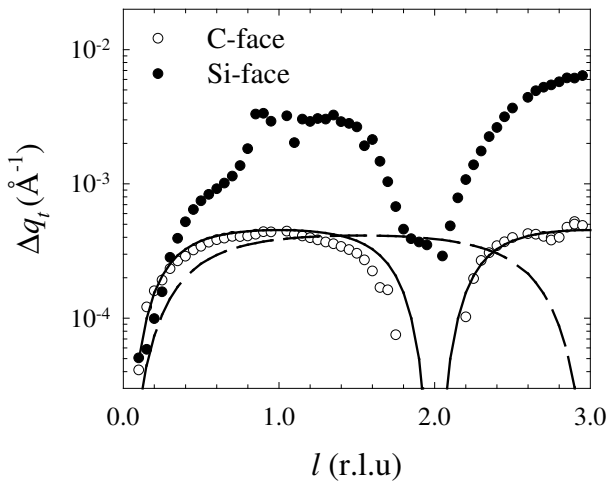}
\end{center}
\vspace{-1em}
\caption{Comparison of the FWHM ($\Delta q_t$) of the
$(00\ell)_{G}$ rod vs. $q_z=\ell c^*_{SiC}$ from ($\bullet$) a
2-layer graphene film grown on the Si-face and ($\circ$) an
8-layer film grown on a C-faces 4H-SiC substrate. Instrument
broadening has been removed for clarification. The lines are fits
to a geometric distribution of steps and step heights with either
the (solid line) 1/2 4H step height or the (dashed line) graphite
step height.} \label{F:spec_rocking}
\end{figure}

Why are the C-face graphite films so much better than the Si-face
films?  While we are not able to explain the details of the growth
mechanism leading to this difference, the x-ray data does point
out a dramatic difference in the surface morphology of the SiC
substrate after film growth. Transverse scans along the specular
$(00\ell)$ rod were taken on both C- and Si-face graphitized
surfaces. The transverse peak widths $\Delta q_t$ are plotted in
Fig.~\ref{F:spec_rocking} versus $\ell$ (SiC units). These scans
reveal a modulation of the width with $\ell$ that is very
different for the C- and Si-face graphitized surfaces. The
peak-width modulation is due to atomic steps.\cite{Lu82} In this
case it is due to steps on the SiC substrate and not steps in the
graphite.  We know this for two reasons.  First, graphite steps
would cause a width modulation of the graphite
$(1,\bar{1},\ell)_G$ rod that is not observed.

Second, the modulation period $\Delta\ell$ is inversely
proportional to the step height: $d_{step}=c_{SiC}/\Delta\ell$.
For both C- and Si-face samples, $\Delta \ell$ corresponds to half
of the 4H unit cell height (5.043$\text{\AA}$) and not to the
graphite step height ($3.337\text{\AA}$) or any multiple.  This is
clearly demonstrated by the fits in Fig.~\ref{F:spec_rocking}. The
fits are to a model of a geometric distributions of steps and step
heights based on either the half 4H unit cell height or the
graphite step height.  It is clear that the graphite steps produce
the wrong period.

The maximum width in Fig.~\ref{F:spec_rocking} is inversely
proportional to the mean distance between steps on the SiC
substrate, $D$, $D=2\pi/\Delta q_t$.\cite{Lu82} Because the
modulation amplitude is much larger for the Si-face sample, we
conclude that the SiC step density is more than an order of
magnitude greater than on the graphitized C-face (we note that the
starting SiC step density before graphitization was nearly the
same for both samples).

Quantitatively, the C-face samples show that the mean terrace
width of the SiC substrate is $D_C\!\sim\! \unit[1.4]{\mu m}$,
while the Si-face samples have a terrace size of $D_{Si}\!\sim\!
\unit[0.2]{\mu m}$. The higher step densities observed after
graphitization on the Si-face substrate correlate with the poorer
long range order of the graphite grown on this face.  Whether this
is a cause or effect relation, remains to be determined.

In conclusion we have shown that ultrathin films of well-ordered
graphite (graphene) can be grown on the SiC(000$\bar{1}$) surface via
Si sublimation. In spite of a small orientational disorder, the long
range order of the graphite on this surface is more than three times
larger than previously reported for graphite growth on the SiC(0001)
surface.

\subsection*{Acknowledgments}
This research was supported by the National Science Foundation under
Grant No. 0404084, by Intel Research, by the Department of Energy
(DE-FG02-02ER45956), and by the Natural Sciences and Engineering
Research Council of Canada (C. A. Jeffrey).  The Advanced Photon Source
is supported by the DOE Office of Basic Energy Sciences, contract
W-31-109-Eng-38.  The $\mu$-CAT beam line is supported through Ames
Lab, operated for the US DOE by Iowa State University under Contract
No.W-7405-Eng-82.  Any opinions, findings, and conclusions or
recommendations expressed herein are those of the authors and do not
necessarily reflect the views of the research sponsors.


\bibliography{SiC-Xray1}

\begin{thebibliography}{14}
\expandafter\ifx\csname natexlab\endcsname\relax\def\natexlab#1{#1}\fi
\expandafter\ifx\csname bibnamefont\endcsname\relax
  \def\bibnamefont#1{#1}\fi
\expandafter\ifx\csname bibfnamefont\endcsname\relax
  \def\bibfnamefont#1{#1}\fi
\expandafter\ifx\csname citenamefont\endcsname\relax
  \def\citenamefont#1{#1}\fi
\expandafter\ifx\csname url\endcsname\relax
  \def\url#1{\texttt{#1}}\fi
\expandafter\ifx\csname urlprefix\endcsname\relax\def\urlprefix{URL }\fi
\providecommand{\bibinfo}[2]{#2}
\providecommand{\eprint}[2][]{\url{#2}}

\bibitem[{\citenamefont{Avouris}(2002)}]{Avouris02}
\bibinfo{author}{\bibfnamefont{{\mbox{See e.g. Ph}}.}~\bibnamefont{Avouris}},
  \bibinfo{journal}{Acc. Chem. Res.} \textbf{\bibinfo{volume}{35}},
  \bibinfo{pages}{1026} (\bibinfo{year}{2002}).

\bibitem[{\citenamefont{Nakada et~al.}(1996)\citenamefont{Nakada, Fujita,
  Dresselhaus, and Dresselhaus}}]{Nakada96}
\bibinfo{author}{\bibfnamefont{K.}~\bibnamefont{Nakada}},
  \bibinfo{author}{\bibfnamefont{M.}~\bibnamefont{Fujita}},
  \bibinfo{author}{\bibfnamefont{G.}~\bibnamefont{Dresselhaus}},
  \bibnamefont{and} \bibinfo{author}{\bibfnamefont{M.~S.}
  \bibnamefont{Dresselhaus}}, \bibinfo{journal}{Phys. Rev. B}
  \textbf{\bibinfo{volume}{54}}, \bibinfo{pages}{17954} (\bibinfo{year}{1996}).

\bibitem[{\citenamefont{Novoselov et~al.}(2005)\citenamefont{Novoselov, Geim,
  Morozov, Jiang, Katsnelson, Grigorieva, Dubonos, and Firsov}}]{Novoselov05b}
\bibinfo{author}{\bibfnamefont{K.~S.} \bibnamefont{Novoselov}},
  \bibinfo{author}{\bibfnamefont{A.~K.} \bibnamefont{Geim}},
  \bibinfo{author}{\bibfnamefont{S.~V.} \bibnamefont{Morozov}},
  \bibinfo{author}{\bibfnamefont{D.}~\bibnamefont{Jiang}},
  \bibinfo{author}{\bibfnamefont{M.~I.} \bibnamefont{Katsnelson}},
  \bibinfo{author}{\bibfnamefont{I.~V.} \bibnamefont{Grigorieva}},
  \bibinfo{author}{\bibfnamefont{S.~V.} \bibnamefont{Dubonos}},
  \bibnamefont{and} \bibinfo{author}{\bibfnamefont{A.~A.}
  \bibnamefont{Firsov}}, \bibinfo{journal}{Nature}
  \textbf{\bibinfo{volume}{438}}, \bibinfo{pages}{197} (\bibinfo{year}{2005}),
  ISSN \bibinfo{issn}{0028-0836},
  \urlprefix\url{http://dx.doi.org/10.1038/nature04233}.

\bibitem[{\citenamefont{Zhang et~al.}(2005)\citenamefont{Zhang, Tan, Stormer,
  and Kim}}]{Zhang05c}
\bibinfo{author}{\bibfnamefont{Y.}~\bibnamefont{Zhang}},
  \bibinfo{author}{\bibfnamefont{Y.-W.} \bibnamefont{Tan}},
  \bibinfo{author}{\bibfnamefont{H.~L.} \bibnamefont{Stormer}},
  \bibnamefont{and} \bibinfo{author}{\bibfnamefont{P.}~\bibnamefont{Kim}},
  \bibinfo{journal}{Nature} \textbf{\bibinfo{volume}{438}},
  \bibinfo{pages}{201} (\bibinfo{year}{2005}), ISSN \bibinfo{issn}{0028-0836},
  \urlprefix\url{http://dx.doi.org/10.1038/nature04235}.

\bibitem[{\citenamefont{Novoselov et~al.}(2004)\citenamefont{Novoselov, Geim,
  Morozov, Jiang, Zhang, Dubonos, Grigorieva, and Firsov}}]{Novoselov04a}
\bibinfo{author}{\bibfnamefont{K.~S.} \bibnamefont{Novoselov}},
  \bibinfo{author}{\bibfnamefont{A.~K.} \bibnamefont{Geim}},
  \bibinfo{author}{\bibfnamefont{S.~V.} \bibnamefont{Morozov}},
  \bibinfo{author}{\bibfnamefont{D.}~\bibnamefont{Jiang}},
  \bibinfo{author}{\bibfnamefont{Y.}~\bibnamefont{Zhang}},
  \bibinfo{author}{\bibfnamefont{S.~V.} \bibnamefont{Dubonos}},
  \bibinfo{author}{\bibfnamefont{I.~V.} \bibnamefont{Grigorieva}},
  \bibnamefont{and} \bibinfo{author}{\bibfnamefont{A.~A.}
  \bibnamefont{Firsov}}, \bibinfo{journal}{Science}
  \textbf{\bibinfo{volume}{306}}, \bibinfo{pages}{666} (\bibinfo{year}{2004}),
  \urlprefix\url{http://www.sciencemag.org/cgi/content/abstract/306/5696/666}.

\bibitem[{\citenamefont{Berger et~al.}(2004)\citenamefont{Berger, Song, Li, Li,
  Ogbazghi, Feng, Dai, Marchenkov, Conrad, First et~al.}}]{Berger04}
\bibinfo{author}{\bibfnamefont{C.}~\bibnamefont{Berger}},
  \bibinfo{author}{\bibfnamefont{Z.}~\bibnamefont{Song}},
  \bibinfo{author}{\bibfnamefont{T.}~\bibnamefont{Li}},
  \bibinfo{author}{\bibfnamefont{X.}~\bibnamefont{Li}},
  \bibinfo{author}{\bibfnamefont{A.~Y.} \bibnamefont{Ogbazghi}},
  \bibinfo{author}{\bibfnamefont{R.}~\bibnamefont{Feng}},
  \bibinfo{author}{\bibfnamefont{Z.}~\bibnamefont{Dai}},
  \bibinfo{author}{\bibfnamefont{A.~N.} \bibnamefont{Marchenkov}},
  \bibinfo{author}{\bibfnamefont{E.~H.} \bibnamefont{Conrad}},
  \bibinfo{author}{\bibfnamefont{P.~N.} \bibnamefont{First}},
  \bibnamefont{et~al.}, \bibinfo{journal}{J. Phys. Chem. B}
  \textbf{\bibinfo{volume}{108}}, \bibinfo{pages}{19912}
  (\bibinfo{year}{2004}).

\bibitem[{\citenamefont{van Bommel et~al.}(1975)\citenamefont{van Bommel,
  Crombeen, and van Tooren}}]{vanBommel75}
\bibinfo{author}{\bibfnamefont{A.~J.} \bibnamefont{van Bommel}},
  \bibinfo{author}{\bibfnamefont{J.~E.} \bibnamefont{Crombeen}},
  \bibnamefont{and} \bibinfo{author}{\bibfnamefont{A.}~\bibnamefont{van
  Tooren}}, \bibinfo{journal}{Surf. Sci.} \textbf{\bibinfo{volume}{48}},
  \bibinfo{pages}{463} (\bibinfo{year}{1975}).

\bibitem[{\citenamefont{Charrier et~al.}(2002)\citenamefont{Charrier, Coati,
  Argunova, Thibaudau, Garreau, Pinchaux, Forbeaux, Debever, Sauvage-Simkin,
  and Themlin}}]{Charrier02}
\bibinfo{author}{\bibfnamefont{A.}~\bibnamefont{Charrier}},
  \bibinfo{author}{\bibfnamefont{A.}~\bibnamefont{Coati}},
  \bibinfo{author}{\bibfnamefont{T.}~\bibnamefont{Argunova}},
  \bibinfo{author}{\bibfnamefont{F.}~\bibnamefont{Thibaudau}},
  \bibinfo{author}{\bibfnamefont{Y.}~\bibnamefont{Garreau}},
  \bibinfo{author}{\bibfnamefont{R.}~\bibnamefont{Pinchaux}},
  \bibinfo{author}{\bibfnamefont{I.}~\bibnamefont{Forbeaux}},
  \bibinfo{author}{\bibfnamefont{J.~M.} \bibnamefont{Debever}},
  \bibinfo{author}{\bibfnamefont{M.}~\bibnamefont{Sauvage-Simkin}},
  \bibnamefont{and} \bibinfo{author}{\bibfnamefont{J.~M.}
  \bibnamefont{Themlin}}, \bibinfo{journal}{J. Appl. Phys.}
  \textbf{\bibinfo{volume}{92}}, \bibinfo{pages}{2479} (\bibinfo{year}{2002}).

\bibitem[{\citenamefont{Forbeaux et~al.}(2000)\citenamefont{Forbeaux, Themlin,
  Charrier, Thibaudau, and Debever}}]{Forbeaux00}
\bibinfo{author}{\bibfnamefont{I.}~\bibnamefont{Forbeaux}},
  \bibinfo{author}{\bibfnamefont{J.~M.} \bibnamefont{Themlin}},
  \bibinfo{author}{\bibfnamefont{A.}~\bibnamefont{Charrier}},
  \bibinfo{author}{\bibfnamefont{F.}~\bibnamefont{Thibaudau}},
  \bibnamefont{and} \bibinfo{author}{\bibfnamefont{J.~M.}
  \bibnamefont{Debever}}, \bibinfo{journal}{Appl. Surf. Sci.}
  \textbf{\bibinfo{volume}{162}}, \bibinfo{pages}{406} (\bibinfo{year}{2000}).

\bibitem[{\citenamefont{Kusunoki et~al.}(2000)\citenamefont{Kusunoki, Suzuki,
  Hirayama, Shibata, and Kaneko}}]{Kusunoki00}
\bibinfo{author}{\bibfnamefont{M.}~\bibnamefont{Kusunoki}},
  \bibinfo{author}{\bibfnamefont{T.}~\bibnamefont{Suzuki}},
  \bibinfo{author}{\bibfnamefont{T.}~\bibnamefont{Hirayama}},
  \bibinfo{author}{\bibfnamefont{N.}~\bibnamefont{Shibata}}, \bibnamefont{and}
  \bibinfo{author}{\bibfnamefont{K.}~\bibnamefont{Kaneko}},
  \bibinfo{journal}{Appl. Phys. Lett.} \textbf{\bibinfo{volume}{77}},
  \bibinfo{pages}{531} (\bibinfo{year}{2000}).

\bibitem[{\citenamefont{Berger et~al.}(2006)\citenamefont{Berger, Song, Li, Wu,
  Brown, Naud, Mayou, Li, Hass, Marchenkov et~al.}}]{Berger06}
\bibinfo{author}{\bibfnamefont{C.}~\bibnamefont{Berger}},
  \bibinfo{author}{\bibfnamefont{Z.}~\bibnamefont{Song}},
  \bibinfo{author}{\bibfnamefont{X.}~\bibnamefont{Li}},
  \bibinfo{author}{\bibfnamefont{X.}~\bibnamefont{Wu}},
  \bibinfo{author}{\bibfnamefont{N.}~\bibnamefont{Brown}},
  \bibinfo{author}{\bibfnamefont{C.}~\bibnamefont{Naud}},
  \bibinfo{author}{\bibfnamefont{D.}~\bibnamefont{Mayou}},
  \bibinfo{author}{\bibfnamefont{T.}~\bibnamefont{Li}},
  \bibinfo{author}{\bibfnamefont{J.}~\bibnamefont{Hass}},
  \bibinfo{author}{\bibfnamefont{A.~N.} \bibnamefont{Marchenkov}},
  \bibnamefont{et~al.}, \bibinfo{journal}{submitted to \emph{Science}}
  (\bibinfo{year}{2006}).

\bibitem[{\citenamefont{{Cree Inc.}}()}]{Cree}
\bibinfo{author}{\bibnamefont{{Cree Inc.}}}, \bibinfo{note}{4600 Silicon Drive,
  Durham, NC 27703}.

\bibitem[{\citenamefont{Li et~al.}(2006)\citenamefont{Li, Ogbazghi, and
  First}}]{Li06}
\bibinfo{author}{\bibfnamefont{T.}~\bibnamefont{Li}},
  \bibinfo{author}{\bibfnamefont{A.~Y.} \bibnamefont{Ogbazghi}},
  \bibnamefont{and} \bibinfo{author}{\bibfnamefont{P.~N.} \bibnamefont{First}},
  \bibinfo{journal}{submitted to \emph{Surface Science}}
  (\bibinfo{year}{2006}).

\bibitem[{\citenamefont{Lu and Lagally}(1982)}]{Lu82}
\bibinfo{author}{\bibfnamefont{T.~M.} \bibnamefont{Lu}} \bibnamefont{and}
  \bibinfo{author}{\bibfnamefont{M.~G.} \bibnamefont{Lagally}},
  \bibinfo{journal}{Surf. Sci.} \textbf{\bibinfo{volume}{120}},
  \bibinfo{pages}{47} (\bibinfo{year}{1982}).

\end{thebibliography}

\end{document}